\documentclass[preprint,aps,superscriptaddress,longbibliography]{revtex4-2}

\usepackage{amssymb}
\usepackage{amsmath}
\usepackage{amsfonts}
\usepackage{bm}
\usepackage{graphicx}
\usepackage{color}

\usepackage[displaymath]{lineno}

\usepackage{hyperref}
\hypersetup{
	colorlinks,
	linkcolor={blue},
	citecolor={blue},
	urlcolor={blue}
}

\newcommand{\CRS}{CeRu$_4$Sn$_6$}
\newcommand{\LRS}{LaRu$_4$Sn$_6$}
\newcommand{\CBP}{Ce$_3$Bi$_4$Pd$_3$}

\begin{document}
%\linenumbers

\begin{center}
{\Large Emergent Topological Semimetal}
\vspace{1cm}

D. M. Kirschbaum$^{1,\ast}$, L. Chen$^{2,\ast}$, D. A. Zocco$^1$, H. Hu$^2$, F. Mazza$^1$, J. Larrea Jim\'{e}nez$^{1,3}$, A. M. Strydom$^4$, D.\ Adroja$^5$, X. Yan$^1$, A. Prokofiev$^1$, Q. Si$^2$, and S. Paschen$^{1,\dagger}$
\vspace{0.5cm}

$^1$Institute of Solid State Physics, Vienna University of Technology, 1040 Vienna, Austria\\
$^2$Department of Physics and Astronomy, Center for Quantum Materials, 6100 Main Street, Rice University, Houston, Texas 77005, USA\\
$^3$Laboratory for Quantum Matter under Extreme Conditions, Institute of Physics, University of S\~{a}o Paulo, 05508-090, Brazil\\
$^4$Physics Department, Highly Correlated Matter Research Group, University of Johannesburg, Auckland Park 2006, South Africa\\
$^5$ISIS Neutron and Muon Source, Science and Technology Facilities Council,
Rutherford Appleton Laboratory, Didcot OX11 0QX, United Kingdom
\vspace{0.5cm}

$^{\dagger}$Corresponding author. E-mail: paschen@ifp.tuwien.ac.at

$^{\ast}$These authors contributed equally to this work.
\end{center}
\vspace{0.5cm}

A material's electronic topology, which is generally described via its Bloch states and the associated bandstructure, will be enriched by the presence of interactions. In metallic settings, the interactions are usually treated through the concept of quasiparticles. Using the genuinely quantum critical heavy fermion compound CeRu$_4$Sn$_6$, we investigate what happens if no well-defined quasiparticles are present. Surprisingly, we discover a topological semimetal phase that emerges from the material's quantum critical state and exhibits a dome structure as a function of magnetic field and pressure. To understand these results, we study a Weyl-Kondo semimetal model at a Kondo destruction quantum critical point. Indeed, it exhibits features in the spectral function that can define topological crossings beyond the quasiparticle picture. We expect our work to stimulate the search for other emergent topological phases.    

\newpage

\noindent Fundamental experimental \cite{Kli80.1,Tsu82.1,Pic97.1,Sam97.1} and theoretical \cite{Tho82.1,Sto99.1,Lau83.1,Aro84.1} discoveries have led to the development of the flourishing field of topological quantum matter \cite{Has10.1,NatMater16.1,Hal17.1,Arm18.1,Has21.1}. A large amount of (noninteracting) topological electronic materials has been predicted \cite{Bra17.1,Ver19.1,Tan19.1,Zha19.1}, and for various materials, topological crossings have been directly visualized by ARPES \cite{Lv19.2,Sob21.1}. Strongly correlated topological phases such as those identified in heavy fermion systems represent a vast new opportunity, both in realizing novel topological phases of matter and in utilizing electron correlations to amplify the topological responses \cite{Che24.1,Pas21.1}. The identification of materials for such strongly correlated topology is inherently harder, given that ab initio calculations of such materials remain a grand challenge and the best ARPES instruments still lack the energy resolution to image details within the extremely narrow heavy fermion bands. One way to proceed is to combine symmetry constraints on the low-energy excitations of strongly correlated electrons with database surveys \cite{Che22.1}. The present work suggests that there may be an alternative route, namely that quantum phase transitions can nucleate emergent topological phases.

The most prominent example of an emergent phase in topologically trivial correlated materials is unconventional superconductivity which, typically, appears as a dome around a quantum critical point (QCP) \cite{Sca12.1,Ste17.2,Pas21.1}. ``Optimum $T_{\rm c}$'', the highest superconducting transition temperature, is reached near the critical tuning parameter value where non-Fermi liquid (NFL) behavior emerges---frequently associated with strange metal characteristics. A thermodynamic argument for the formation of emergent phases is that the entropy accumulated at a QCP is released as a new phase forms \cite{Pas21.1}. It is striking that this phenomenon is seen across many different materials classes, now incorporating also moir\'e heterostructures and other flat-band systems \cite{Che24.1}. In most cases, the NFL behavior extends to much larger temperatures and, at high temperatures, into a much wider tuning parameter range than the superconducting dome. As such, the ``fan'' of NFL behavior can be used as a pointer to the emergent phase.

Metallic heavy fermion compounds are a well-established platform for realizing such QCPs and emergent superconductivity \cite{Pas21.1}. Kondo insulators---gapped representatives of heavy fermion compounds---have larger energy scales than heavy fermion metals, and larger control parameter changes are needed to tune them from one phase to another \cite{Si13.1}. The intermediate situation---a semimetallic heavy fermion state---is of great interest in the context of topological semimetals. We are not aware of any experiments that have investigated its potential connection to quantum criticality. Here we find that a Weyl--Kondo semimetal phase emerges from the quantum critical state of the heavy fermion semimetal \CRS, with a dome-like shape as a function of pressure and magnetic field. We propose that this situation may not be unique to \CRS, and amounts to a new design principle for the discovery of other correlation-driven topological phases.

\begin{figure}[h]
    \centering
    \includegraphics[width=\textwidth]{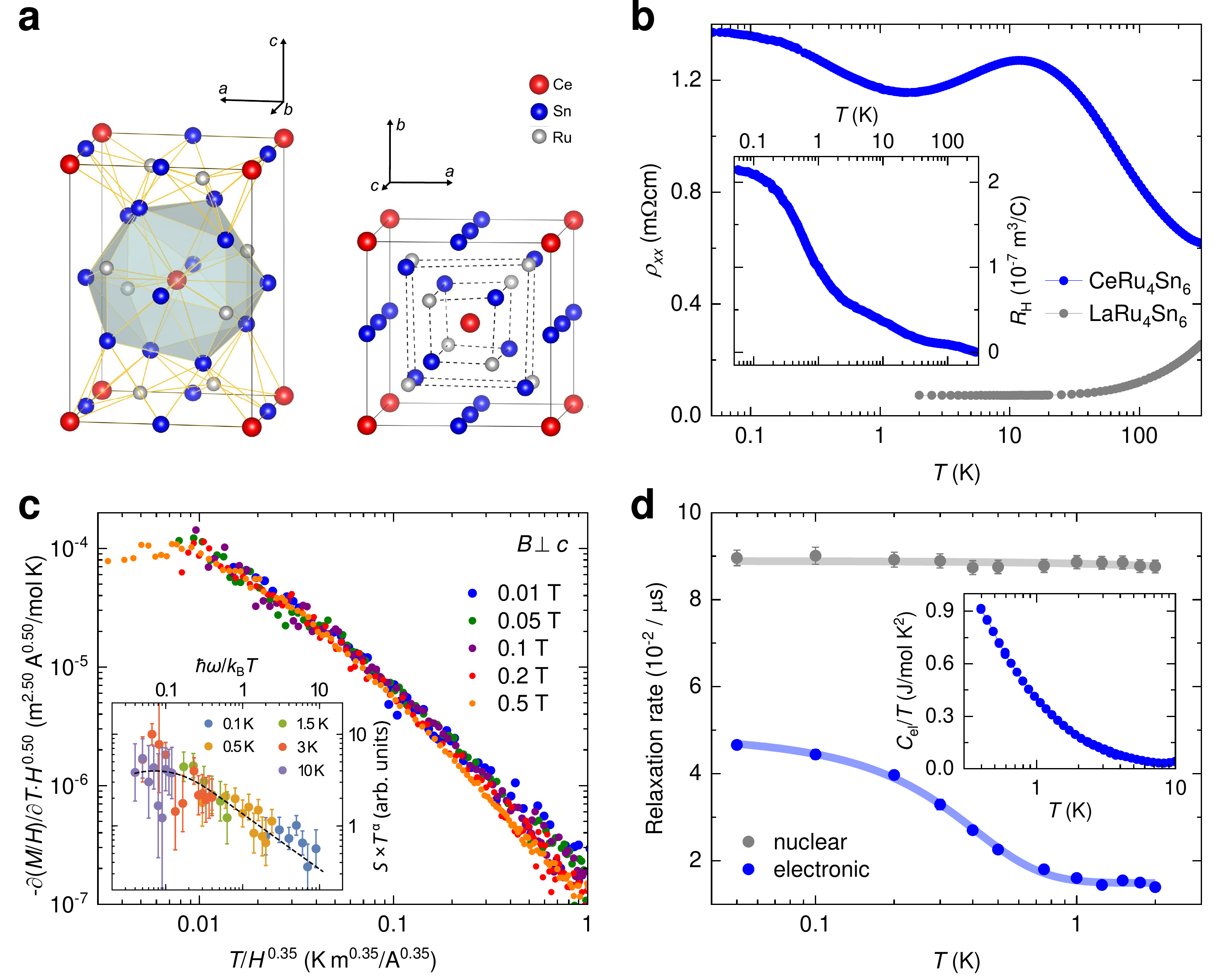}

    \caption{{\bf Overview and characterization of \CRS.} {\bf a}, Unit cell (space group 121, $I\bar{4}2m$) viewed nearly along the tetragonal $b$ ($=a$) direction, with the polyhedron of nearest neighbors around the central Ce atom (left), and nearly along the $c$ direction (right), where the broken inversion symmetry is evident. {\bf b}, Temperature-dependent electrical resistivity (with $j \parallel c$) of \CRS\ and the nonmagnetic reference compound \LRS\ at zero field (main panel), and the Hall coefficient $R_\mathrm{H}$ of \CRS, which saturates at low temperatures to a finite carrier concentration (inset), as expected for a semimetal. {\bf c}, Signatures of pristine quantum criticality in the thermodynamic (main panel) and inelastic neutron scattering response (inset) of \CRS. Figure adapted from Ref.\cite{Fuh21.1}. {\bf d}, Temperature dependence of the electronic (blue) and nuclear contribution (grey) to the muon spin relaxation rate obtained from ZF-$\mu$SR measurements, indicating the absence of magnetic order (Supplementary Information section A). The electronic specific heat coefficient $C_{\rm el}/T$ increases smoothly with decreasing temperature (inset), as expected for quantum criticality.}
    \label{fig1}
\end{figure}

\CRS\ crystallizes in the noncentrosymmetric tetragonal YRu$_4$Sn$_6$-type structure of space group 121 ($I\bar{4}2m$) \cite{Poe97.1} (Fig.\,\ref{fig1}a). Together with the heavy constituting elements (atomic masses between 100 and 140) and a corresponding large atomic spin-orbit coupling, this may promote topological band crossings. Indeed, within the density functional theory plus Gutzwiller scheme, Weyl nodes were predicted to exist near the Fermi energy \cite{Xu17.1}. The temperature-dependent electrical resistivity $\rho(T)$ of \CRS\ is typical of a semimetal, with a modest increase of $\rho$ with decreasing temperature, whereas the non-$f$ reference compound \LRS\ is a simple metal (Fig.\,\ref{fig1}b). This indicates that the Kondo interaction initiates the opening of a (pseudo)gap, though a full gap does not form. The saturation of the Hall coefficient $R_{\rm H}$ to a small but finite value (0.014 carriers per unit cell in a single band picture) at the lowest temperatures (inset) supports the classification of \CRS\ as a semimetal. In the first established Weyl-Kondo semimetal \CBP\ \cite{Dzs17.1,Dzs21.1}, it was indeed concluded that Weyl nodes, situated within an otherwise Kondo insulating background \cite{Dzs22.1}, prevent a full gap from opening. This feature also appears in Weyl-Kondo semimetal models \cite{Lai18.1,Gre20.1}.

However, scaling collapses of previous magnetization and inelastic neutron scattering data (Fig.\,\ref{fig1}c) revealed genuine quantum criticality of beyond Landau's order-parameter-fluctuation type \cite{Fuh21.1}, as expected for a Kondo destruction QCP \cite{Si01.1}, seemingly excluding the presence of a topological phase. The scenario of quantum criticality is also consistent with zero-field muon spin rotation ($\mu$SR) and specific heat data (Fig.\,\ref{fig1}d): both increase smoothly with decreasing temperature, evidencing enhanced fluctuations, and lack signatures of ordering (Supplementary Information section A). All these set the stage for what we report next, namely our discovery of Weyl-Kondo semimetal signatures that emerge out of this quantum critical state, showing that the two phenomena are in fact compatible with each other. 

\begin{figure}[t!]
    \centering
    \includegraphics[width=\textwidth]{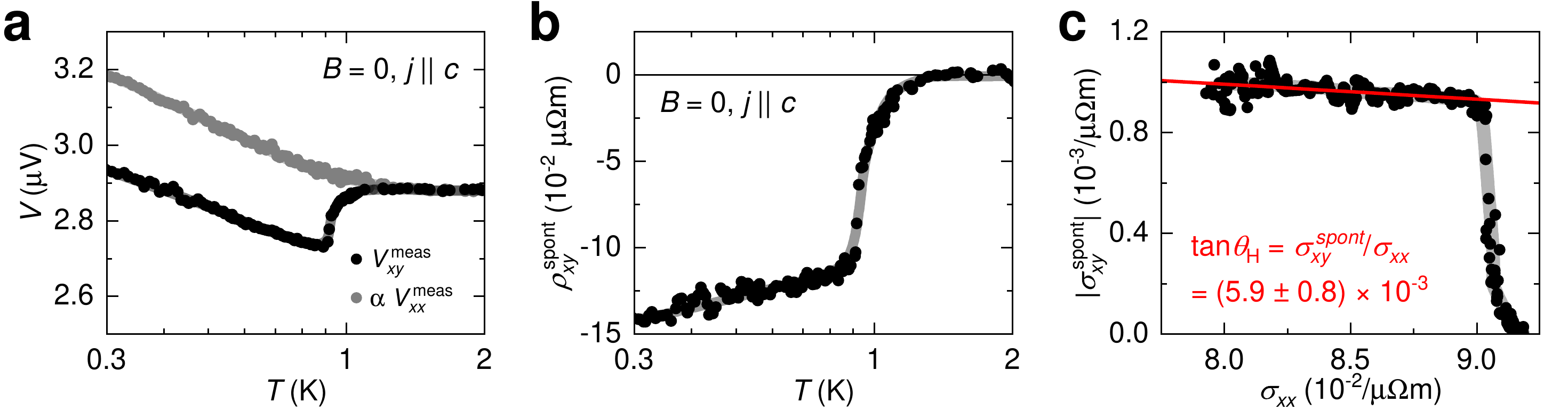}

    \caption{{\bf Spontaneous Hall effect in \CRS.} {\bf a}, Temperature-dependent voltage measured on the Hall contacts $V_{xy}^\mathrm{meas}$ (black) and on the resistivity contacts $V_{xx}^\mathrm{meas}$ (grey), with the latter scaled by the factor $\alpha$ = 0.134 to $V_{xy}^\mathrm{meas}$ above the onset of the anomaly, all in zero magnetic field. Note that the anomaly is visible only in $V_{xy}^\mathrm{meas}$, indicating that it is a transverse response. The curves were measured at ambient pressure (1\,bar) and are exemplary for the analog behavior observed at higher pressures. {\bf b}, Zero-field Hall resistivity as a function of temperature, indicating the onset of a spontaneous Hall effect below 1\,K. The curve was obtained from the difference $R_{xy} = R_{xy}^\mathrm{meas} - \alpha R_{xx}$ (Supplementary Information section B). {\bf c}, Spontaneous Hall conductivity $\sigma_{xy}^{\rm spont}$ as a function of the longitudinal conductivity $\sigma_{xx}$, with temperature $T$ as implicit parameter. Below 0.9\,K, $\sigma_{xy}^{\rm spont}$ is linear in $\sigma_{xx}$. The red line corresponds to a linear fit in this range and gives an estimate of the Hall angle.}
    \label{fig2}
\end{figure}

A transverse voltage contribution appears in \CRS\ below 1\,K, in the presence of a longitudinal electric field but in the absence of any applied magnetic field; it has no correspondence in the longitudinal channel (Fig.\,\ref{fig2}a). As \CRS\ is nonmagnetic, this signal is identified as the spontaneous (nonlinear) Hall effect theoretically expected for a time-reversal invariant but inversion symmetry broken Weyl semimetal \cite{Sod15.1} and first observed in \CBP\ \cite{Dzs21.1}. The corresponding temperature-dependent spontaneous (nonlinear) Hall resistivity $\rho_{xy}^{\rm spont}(T)$ is shown in Fig.\,\ref{fig2}b (see Supplementary Information section B for contact misalignment corrections). Plotting the spontaneous (nonlinear) Hall conductivity $\sigma_{xy}^{\rm spont}$ vs the longitudinal conductivity $\sigma_{xx}$, with temperature as an implicit parameter, reveals a linear relationship below 1\,K. This points to the intrinsic nature of the Hall response \cite{Du19.1}. The slope $\sigma_{xy}^{\rm spont}/\sigma_{xx} \approx 6\times 10^{-3}$, which corresponds to the Hall angle $\tan \theta_\mathrm{H}$, is 60 times larger than the maximum value theoretically expected for TaAs (when the Fermi energy is artificially moved to the energy of the Weyl nodes) \cite{Zha18.1}. Yet, it is distinctly smaller than the giant response found for \CBP\ \cite{Dzs21.1}. One might be tempted to conclude that quantum criticality weakens the Weyl-Kondo semimetal state and that the two phenomena compete. Our measurements under pressure and magnetic field reveal this is not the case: Instead, the phase discovered here is a new ``emergent'' Weyl-Kondo semimetal, stabilized by quantum critical fluctuations.

\begin{figure}[t!]
\vspace{-1cm}

    \centering
    \includegraphics[width=\textwidth]{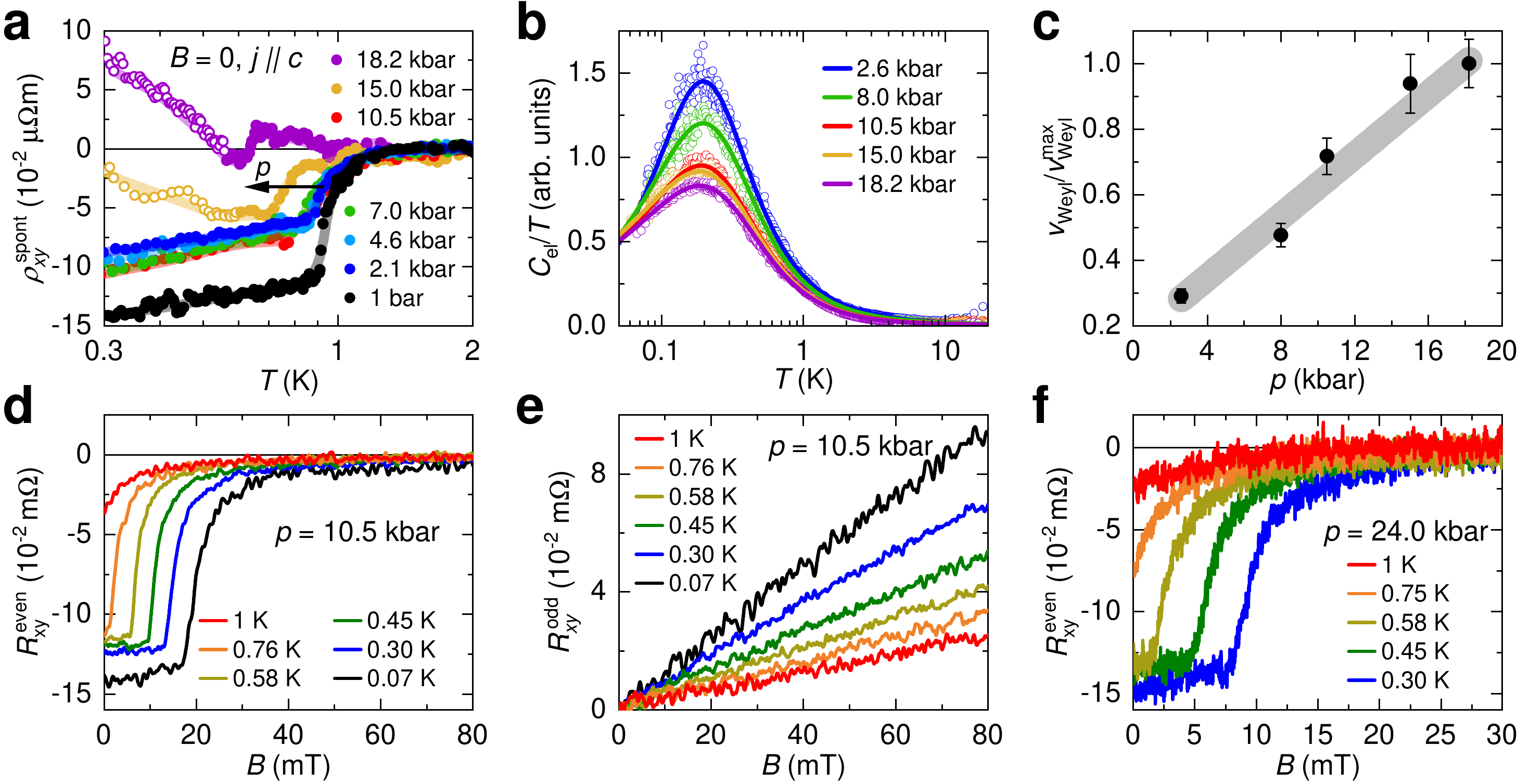}

    \caption{{\bf Pressure and magnetic field tuning.} {\bf a}, Spontaneous Hall resistivity $\rho_{xy}^{\rm spont}$ as a function of temperature for different pressures $p$, showing the suppression of the onset temperature $T_\mathrm{H}$ with increasing $p$. Above 15\,kbar, the background subtraction via $\alpha R_{xx}$ introduces sizable error due to the steep increase of $R_{xx}(T)$ at the lowest temperatures (Supplementary Information section E; affected parts of the curves are shown with open symbols). $\sigma_{xy}^\mathrm{spont}(T)$ is less sensitive to the $R_{xx}$ background and is almost fully suppressed at 18.2\,kbar (Supplementary Information section B). {\bf b}, Electronic specific heat $C_\mathrm{el}/T$ vs $T$ for different pressures obtained via AC calorimetry. The solid lines are phenomenological fits describing a crossover between non-Fermi liquid behavior and a contribution from linearly dispersing Weyl bands (Supplementary Information section C). {\bf c}, Pressure-dependent Weyl velocity $v_\mathrm{Weyl}$ normalized to its maximum value. It was calculated from the Weyl contribution obtained from the fits in panel b (Supplemental Information section C). The error bars represent standard deviations of the fit.  The increase of $v_\mathrm{Weyl}$ and the overall suppression of $C_\mathrm{el}$ with pressure indicate that pressure tunes the system away from the quantum critical point. {\bf d},\,{\bf e}, As expected for the finite-field extension of the spontaneous Hall effect, the magnetic field dependent Hall resistance $R_{xy}(B)$ can be decomposed in an even-in-field component $R_{xy}^\mathrm{even}$ (panel d) and an odd-in-field component $R_{xy}^\mathrm{odd}$ (panel e) (Supplementary Information section\,D). $R_{xy}^\mathrm{even}$ is suppressed with increasing $T$ and $B$. {\bf f}, The even-in-field Hall component $R_{xy}^\mathrm{even}(B)$ persists even up to the highest pressure of 24\,kbar reached in these experiments, where detection of the zero-field spontaneous Hall effect was no longer possible due to the rapid increase of $R_{xx}(T)$ in the relevant temperature range. Also here, the onset field $B_\mathrm{H}^\mathrm{even}$ is suppressed with increasing $p$, in agreement with the suppression of $T_\mathrm{H}$ presented in panel\,a.} 
    \label{fig3}
\end{figure}

\begin{figure}[t!]
    \centering
    \includegraphics[width=\textwidth]{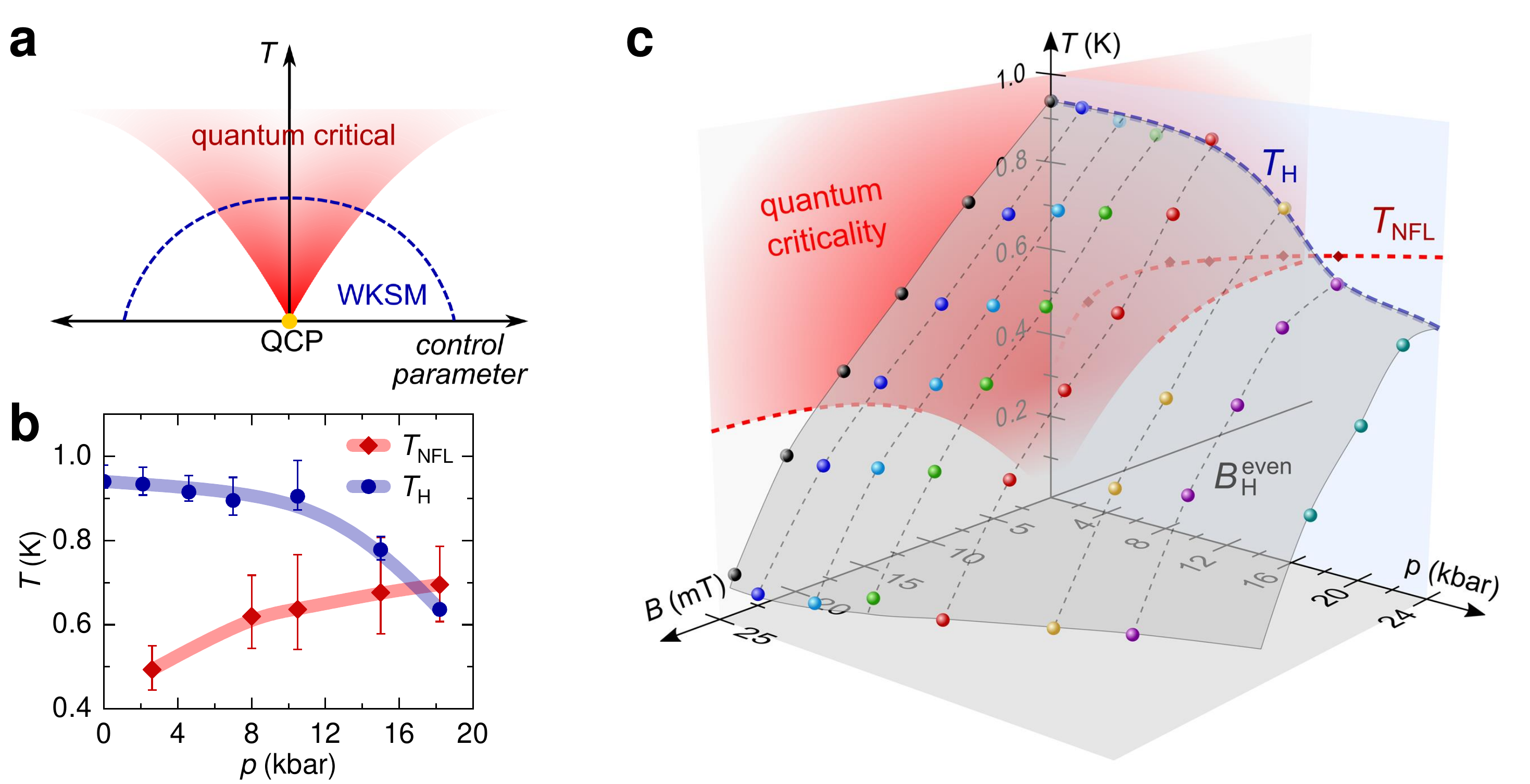}

    \caption{{\bf Pressure and magnetic field tuned phase diagram of \CRS.} {\bf a}, Cartoon of the phase diagram expected for an emergent Weyl-Kondo semimetal, created by quantum criticality of the beyond order-parameter-fluctuation type. {\bf b}, Temperature-pressure phase diagram of \CRS\ at zero magnetic field, defined by the onset temperature $T_\mathrm{H}$ of the spontaneous Hall effect (orange dots) and the lower bound of the quantum critical fan $T_\mathrm{NFL}$ (blue diamonds) obtained from fits to the specific heat data (Fig.\,\ref{fig3}b). The results are consistent with the expected behavior sketched in panel a. {\bf c}, Temperature-pressure-magnetic field phase diagram with the onset temperatures $T_\mathrm{H}$ (in the $B=0$ plane) and fields $B_{xy}^\mathrm{even}$ of the spontaneous Hall effect. The points were obtained from the zero-field $\rho_{xy}^{\rm spont}(T)$ curves shown in Fig.\,\hyperref[Fig3]{\ref{fig3}a} and the even-in-field contribution to the Hall isotherms shown exemplarily in Fig.\,\hyperref[Fig3]{\ref{fig3}d,f} (Supplementary Information sections B and D). These signatures, associated with Weyl-Kondo physics, are suppressed with both pressure and magnetic field and thus form a dome of Weyl-Kondo semimetal behavior centered around the quantum critical point (QCP). The red shading of the quantum critical fan in the $p=0$ plane reflects the temperature range in which the scaling collapse of the inelastic neutron scattering holds, and the dashed boundary was derived from the NFL behavior in the magnetization $M(T)$ curves at different fields \cite{Fuh21.1}. The red dashed line and diamonds in the $B=0$ plane correspond to $T_\mathrm{NFL}$ in panel b.} 
    \label{fig4}
\end{figure}

We first discuss the measurements under hydrostatic pressure. We observe that both the magnitude of $\rho_{xy}^{\rm spont}$ and its onset temperature decrease with increasing pressure (Fig.\,\ref{fig3}a). From the pressure-dependent specific heat measurements (Fig.\,\ref{fig3}b) we conclude that pressure suppresses quantum criticality and that, thus, the spontaneous Hall signal is strongest at ``optimum'' quantum criticality. In detail, to disentangle the Weyl-Kondo semimetal and quantum criticality contributions to the specific heat, we fit the data with a phenomenological function that describes a crossover between the two contributions (Supplementary Information section C). The lower bound $T_{\rm NFL}$ of the quantum critical fan extracted from this analysis increases with increasing pressure (Fig.\,\ref{fig4}b), providing clear evidence that pressure drives the system away from quantum criticality. The Weyl velocity $v_{\rm Weyl}$, calculated from the prefactor $\Gamma$ of the Weyl-Kondo semimetal contribution $C_{\rm el} = \Gamma T^3$ \cite{Dzs17.1,Lai18.1}, also increases with increasing pressure (Fig.\,\ref{fig3}c). As $v_{\rm Weyl}$ is the slope of the Weyl dispersion, the smallest value corresponds to the flattest Weyl band and thus to the most strongly correlated state. Maximal correlation strength at quantum criticality is well established for topologically trivial heavy fermion compounds and related materials \cite{Che24.1,Pas21.1}, but has not been demonstrated for correlation-driven topological semimetals.

Next, we describe our magnetic field tuning experiments. As shown previously \cite{Fuh21.1}, the quantum criticality of \CRS\ is suppressed by magnetic fields \cite{Fuh21.1} (see also Fig.\,\ref{fig1}c). The Berry curvature probing quantity in finite fields is the even-in-field Hall resistance $R_{xy}^{\rm even}$ \cite{Dzs21.1} (Supplementary Information section D). Starting in the Weyl-Kondo semimetal state, a finite magnetic field suppresses this even-in-field contribution (Fig.\,\ref{fig3}d,\,f for exemplary iso-bars), whereas the (normal) odd-in-field Hall resistance $R_{xy}^\mathrm{odd}$ is featureless (Fig.\,\ref{fig3}e). Note that only tiny fields are needed to suppress the Weyl response, in contrast to the sizable fields needed for Weyl node annihilation in \CBP\ \cite{Dzs22.1} under Zeeman coupling tuning \cite{Gre20.1x}. This confirms the result from pressure tuning that the Weyl-Kondo semimetal is strongest at ambient conditions ($p=B=0$), where it is quantum critical.

By defining Weyl-Kondo semimetal onset temperatures $T_{\rm H}$ and fields $B_{\rm H}^{\rm even}$ (Supplementary Information section D) we can construct a temperature--pressure--magnetic field phase diagram of \CRS\ (Fig.\,\ref{fig4}c). It shows a dome of Weyl-Kondo semimetal behavior centered around the QCP, similar to the cartoon in Fig.\,\ref{fig4}a, indicating that quantum critical fluctuations stabilize the Weyl-Kondo semimetal as an emergent topological phase. The understanding of this experimental observation poses a challenge. The definition of band topology roots in the particle (or quasiparticle) description of electrons in solids, via density functional theory in the simplest case or through effective renormalized band descriptions. At a Kondo destruction QCP, as evidenced in \CRS, quasiparticles appear to be absent \cite{Che23.1}. How can a quantum critical state that loses quasiparticles nucleate a topological semimetal?

To address this question, we begin by recognizing that the loss of quasiparticles invalidates the standard definition of topology in terms of Bloch states and the associated Berry curvature. Instead, it has recently been shown that topological nodes can be formulated as crossings of the single-particle spectral functions \cite{Hu21.1x}. The key to this formulation is that the eigenfunctions of the single-particle Green's function in interacting systems form a representation of the space group, as Bloch functions do in the absence of interactions. Even for single-particle excitations that break the quasiparticle description, symmetry constraints ensure that the single-particle spectral functions, which are eigenvalues of the Green's function, cross at the symmetry-dictated wavevectors \cite{Hu21.1x}. This crossing corresponds to a topological node: A frequency-dependent Berry curvature, defined in terms of the Green's function eigenfunctions, specifies that the Berry flux surrounding such a node is quantized \cite{Set23.2x}.

\begin{figure}[t!]
    \centering
    \includegraphics[width=0.9\textwidth]{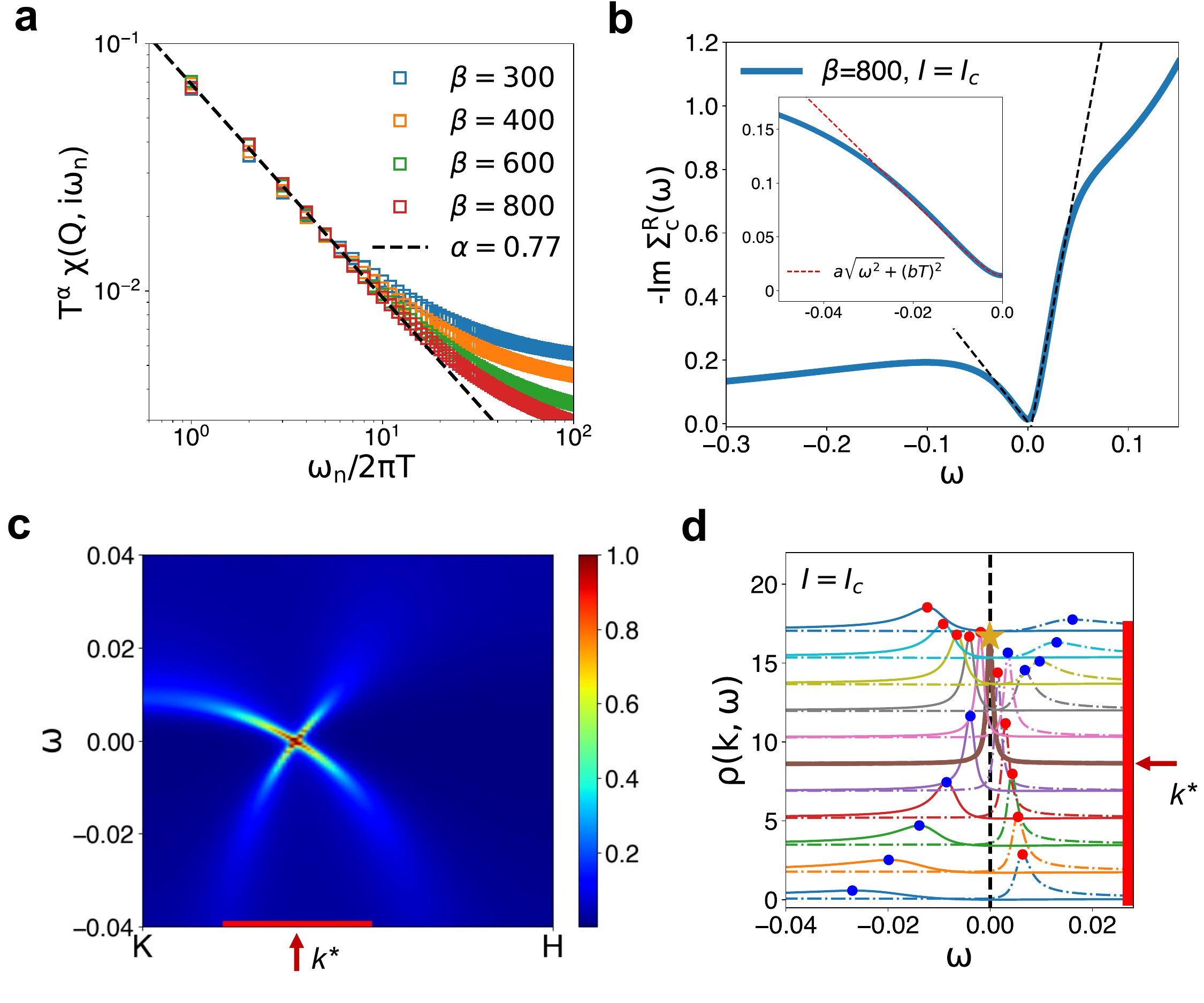}
    \caption{{\bf Kondo destruction quantum criticality nucleating a Weyl-Kondo semimetal.} {\bf a},  Kondo destruction QCP in an Anderson lattice model, defined on a 3D kagome lattice \cite{Hu21.1x}, as signified by an $\omega/T$ scaling of the dynamical lattice spin susceptibility. (Throughout this figure, we set $\hbar=k_{\rm B}=1$.) {\bf b}, The imaginary part of the conduction electron self-energy at the Kondo destruction QCP in the real frequency domain, obtained from a Pad\'e decomposition. In the low-frequency region, it follows a linear-in-frequency dependence (main plot). Taking into account the small but nonzero temperature $T$, it is well-described by $-\Im \Sigma_c(T, \omega) \approx a\sqrt{\omega^2 + (bT)^2}$ at low frequencies, with the dimensionless fitting parameters $a=4.09$ and $b=2.74$ (inset). {\bf c}, Spectral function of the $f$ electrons plotted along the high symmetry $K-H$ line of the Brillouin zone, where a spectral crossing is identified. {\bf d}, Momentum-resolved energy distribution curves, with the red and blue dots highlighting the maxima and the corresponding dispersion. The red bars in c and d indicate the same range of wave vectors. At $k^{\ast}$, the two branches overlap (maximum highlighted by beige star), with the corresponding Weyl-nodal crossing of the spectral functions marked by the thick brown curve.}
    \label{fig5}
\end{figure}

To demonstrate this effect near a QCP, we study an Anderson-lattice model whose underlying (noninteracting) bandstructure contains symmetry-protected nodes (Supplementary Information section F). This model features a competition between the Ruderman-Kittel-Kasuya-Yosida (RKKY) and Kondo interactions, which we study in terms of an extended dynamical mean-field theory \cite{Hu22.1x}. It realizes a Kondo destruction QCP \cite{Si01.1}, where the Landau quasiparticles are destroyed. At this QCP, $k_{\rm B}T$ emerges as the only energy scale in the system, giving rise to dynamical properties that obey a scaling in terms of the ratio of $\hbar\omega$ to $k_{\rm B}T$. We illustrate this by calculating the dynamical spin susceptibility at the wavevector where it peaks, finding it to have the dynamical scaling form with a fractional exponent ($0.77$) as shown in Fig.\,\ref{fig5}a. The quantum criticality is also captured by the single-particle excitations. Fig.\,\ref{fig5}b shows the conduction electron self-energy in real frequency at the Kondo destruction QCP, obtained from an analytical continuation (Supplementary Information section F). In the relevant low-energy regime, the imaginary part of the self-energy ($-\Im \Sigma_{c}^{R}(\omega)$) exhibits a linear-in-$\omega$ dependence, which implies that the quasiparticle residue $z$ vanishes or, in other words, the loss of quasiparticles.

To see how this quantum criticality nucleates topology, we show in Fig.\,\ref{fig5}c the spectral function of the $f$ electron along a high-symmetry line $K-H$ in the Brillouin zone (Supplementary Information section F). Due to the symmetry constraint, the peaks of the spectral function intersect, resulting in a non-Fermi liquid form of a Weyl point. We then examine the wavevector region marked by the red bar (Fig.\,\ref{fig5}c) and plot the momentum-resolved energy distribution curve, which is shown in Fig.\,\ref{fig5}d. The red and blue dots track the peaks in the frequency dependence of the spectral functions, showing a Weyl crossing close to the Fermi energy. 

Our work has advanced a new design principle for correlation-driven topological phases, namely that a topological semimetal can emerge out of quantum criticality beyond the Landau order-parameter-fluctuation type. Both our experimental and theoretical results show that the signatures of the quantum criticality persist into the emergent Weyl-Kondo semimetal phase. This differs from the situation of emergent superconductivity, where the phase transition to the superconducting state releases the entropy of the quantum critical fluctuations and restores the quasiparticles of a (superconducting) Fermi liquid. Topological semimetals are not defined by a (Landau) order parameter but by topological indices and are therefore not bound by a thermal phase transition. Consequently, the Weyl-Kondo semimetal itself is expected to be a non-Fermi liquid. This could be tested by future experiments, for instance by measuring whether shot noise is reduced from its Fermi-liquid value \cite{Che23.1}. It would also be illuminating to test whether the emergent topological semimetal is characterized by enhanced entanglement.

\newpage

%%%%%%%%% FIGURE 1 %%%%%%%%%%%%%%%%%%%%%%%%%%%%%%%%%%%%%%%%%%%%%%%%%%%%%%%%%%%%%%%%

%%%%%%%%% FIGURE 2 %%%%%%%%%%%%%%%%%%%%%%%%%%%%%%%%%%%%%%%%%%%%%%%%%%%%%%%%%%%%%%%%

%%%%%%%%% FIGURE 3 %%%%%%%%%%%%%%%%%%%%%%%%%%%%%%%%%%%%%%%%%%%%%%%%%%%%%%%%%%%%%%%%

%%%%%%%%% FIGURE 4 %%%%%%%%%%%%%%%%%%%%%%%%%%%%%%%%%%%%%%%%%%%%%%%%%%%%%%%%%%%%%%%%

%%%%%%%%% FIGURE 5 %%%%%%%%%%%%%%%%%%%%%%%%%%%%%%%%%%%%%%%%%%%%%%%%%%%%%%%%%%%%%%%%

\clearpage
\newpage

{\noindent\large\bf Acknowledgements}\\
We thank Valentina Martelli and Monika Lu\v{z}nik for support during the low-temperature pressure experiments, Adrian Hillier for help with $\mu$SR data extraction and advice on data analysis, and Ernst Bauer, Jennifer Cano, Chandan Setty, Shouvik Sur, Mathieu Taupin, Maia Vergniory, and Fang Xie for fruitful discussions. The work in Vienna was supported by the Austrian Science Fund (FWF grants I4047, SFB F86 ``Q-M\&S'', I5868-N/FOR5249 ``QUAST'', and 10.55776/COE1 ``quantA''), the European Microkelvin Platform (H2020 project 824109), and the European Research Council (ERC Advanced Grant 101055088-CorMeTop). The work at Rice has primarily been supported by the Air Force Office of Scientific Research under Grant No. FA9550-21-1-0356 (L.C.), by the National Science Foundation under Grant No. DMR-2220603 (H.H.), by the Robert A. Welch Foundation Grant No. C-1411 and the Vannevar Bush Faculty Fellowship ONR-VB N00014-23-1-2870 (Q.S.). The majority of the computational calculations have been performed on the Shared University Grid at Rice funded by NSF under Grant EIA-0216467, a partnership between Rice University, Sun Microsystems, and Sigma Solutions, Inc., the Big-Data Private-Cloud Research Cyberinfrastructure MRI-award funded by NSF under Grant No. CNS-1338099, and the Extreme Science and Engineering Discovery Environment (XSEDE) by NSF under Grant No. DMR170109. J.L.J.\ acknowledges the JP FAPESP 2018/08845-3 and the CNPq 310065/2021-6. A.M.S.\ thanks the SA-NRF (93549) and the URC/FRC of UJ for financial assistance.

%\bibliographystyle{naturemagallauthors} 
%%\bibliography{silke_all,diana}
%\bibliography{../../silke_all}

\end{document}